\documentclass[preprint, aps]{revtex4-1}

\begin{document}
\title{Quantum Mechanics from Newton's Second Law and the Canonical
Commutation Relation $[X,P]=i$}
\author{Mark C. Palenik}

\affiliation{Department of Physics, Purdue University, West Lafayette, IN,
47909, USA}

\begin{abstract}
Despite the fact that it has been known since the time of Heisenberg that
quantum operators obey a quantum version of Newton's laws, students are often
told that derivations of quantum mechanics must necessarily follow from the
Hamiltonian or Lagrangian formulations of mechanics.  Here, we first derive the
existing Heisenberg equations of motion from Newton's laws and the uncertainty
principle using only the equations $F=\frac{dP}{dt}$, $P=m\frac{dV}{dt}$, and
$\left[X,P\right]=i$.  Then, a new expression for the propagator is derived that
makes a connection between time evolution in quantum mechanics and the motion of
a classical particle under Newton's laws.  The propagator is solved for three
cases where an exact solution is possible 1) the free particle 2) the harmonic
oscillator 3) a constant force, or linear potential in the standard
interpretation. We then show that for a general force F(X), by
Taylor expanding X(t) in time, we can use this methodology to reproduce the
Feynman path integral formula for the propagator.  Such a picture may be useful
for students as they make the transition from classical to quantum mechanics and
help solidify the equivalence of the Hamiltonian, Lagrangian, and Newtonian
formulations of physics in their minds.
\end{abstract}

\maketitle

\section{Introduction}

Typical introductory quantum mechanics classes take place after students have
studied, at least to some extent, the Hamiltonian and Lagrangian formulations of
classical mechanics.  The role of the Hamiltonian and the Schr\"{o}dinger
equation are emphasized, and it is often taught that these energy-based
formulations of physics are more general because they allow physics to be
extended into the quantum regime.  Quantum mechanics is, then, treated as a
theory that depends on the existence of Lagrangian and Hamiltonian mechanics
and where Newton's laws no longer have any applicability, outside of the
occasional reference to the Ehrenfest theorem\cite{ehrenfest}.  This treatment
is apparent from the current standard introductory quantum mechanics
textbooks\cite{weinbergQM,griffithsQM,SakuraiQM}.

Heisenberg, in his initial formulation of matrix mechanics, made use of
correspondence between the time evolution of quantum operators and classical
particles\cite{AJPHeisenberg}.  And while quantum and classical correspondence
has been acknowledged since the earliest days of quantum
physics\cite{VVCorrespondence}, it seems that the Newtonian-like dynamics of
quantum operators has never been used as a starting point for the development of
quantum physics.  The Hamiltonian and in more advanced courses, Lagrangian
formulation of Feynman\cite{PathIntegral}, are generally taken to be both
necessary and fundamental.

We will first rederive Heisenberg picture mechanics starting from Newton's laws
plus the uncertainty principle.  This is presented mainly as a tool for
reinforcing the equivalence between the Newtonian and Hamiltonian formulations
of physics, even within the quantum regime. On its own, however, it does not
clearly formulate the utility of quasi-Newtonian principles in quantum physics.

A Newton-like formulation of quantum mechanics is possible, which we demonstrate
through the derivation of a new expression for the propagator.  This expression
utilizes the concept of a position operator that evolves in time in an analogous
manner to the position of a Newtonian particle.  The propagator is then solved
for three cases where an exact solution is possible: the free particle, a
harmonic oscillator, and a constant force.

Our expression emphasizes the time-evolution of the operator $X(t)$, just as in
classical mechanics, the classical variable $X(t)$ evolves according to Newton's
laws.  The initial value $X_0$ and subsequent derivatives $\frac{1}{m}P$ and
$\frac{1}{m}F$ are used to build the time dependence of $X(t)$ without
referencing the Hamiltonian or any energy-based formulation of mechanics.
Although there have been descriptions of quantum mechanics that treat it as a
classical theory with random Newtonian forces leading to a stochastic
differential equation\cite{StochasticQM,SchrodingerFromNewton}, a
Newtonian-based derivation of standard quantum physics does not appear to have
been previously developed.

\section{Reproducing the Heisenberg Equations of Motion}

We will start by reproducing Heisenberg picture quantum mechanics, defined by
the relation (in units where $\hbar=1$)
\begin{equation}
i[H,O] = \frac{\partial}{\partial t}O
\end{equation}
from the equations
\begin{eqnarray}
F&=&\frac{dP}{dt}\\
P&=m&\frac{dX}{dt}\\
\left[X,P\right]&=&i
\end{eqnarray}

We can begin by finding the commutator of $[X^n,P]$ for positive n. Using the
third equation, we can rewrite the commutator as:
\begin{eqnarray}
\nonumber
X^nP - PX^n &=& X^nP - X^{n-1}PX + X^{n-1}PX - X^{n-2}PX^2 + \ldots - PX^n\\
\nonumber
&=&X^{n-1}[X,P] + X^{n-2}[X,P]X +\ldots+[X,P]X^{n-1}\\
&=&i nX^{n-1} = i\frac{d}{dX}X^n
\label{CommDerivX1}
\end{eqnarray}

For negative powers of $X$, we can write
\begin{eqnarray}
\nonumber
[X^{-n},P] &=& X^{-n}[P,X^{n}]X^{-n}\\
\nonumber
&=& -iX^{-n}nX^{n-1}X^{-n}\\
&=& -inX^{-n-1} = i\frac{d}{dX}X^{-n}
\label{CommDerivX2} 
\end{eqnarray}
and in either case, it is clear that commuting a power of $X$ with $P$ results
in its derivative with respect to $X$.

Starting with some arbitrary function of X, $O(X)$, it can be Laurent expanded
as:
\begin{equation}
O(X)=\sum_{n=-\infty}^{\infty} C_nX^n
\end{equation}
where the $C_n$'s are constants.

From equations \ref{CommDerivX1} and \ref{CommDerivX2}, the commutator of P with
each term in the Laurent series results in the derivative of that term with
respect to X.  Thus:
\begin{equation}
[O(X),P]=i\frac{d}{dX}O(X)
\end{equation}

The same argument can be used to show that for a function of momentum $O(P)$
\begin{equation}
[O(P),X] = -i\frac{d}{dP}O(P)
\end{equation}

The Laurent expansion of $O$ also provides a convenient representation in which
to find the time derivative of $O$.  Since in quantum mechanics, the commutators
$\left[X,\frac{dX}{dt}\right]$ and $\left[P,\frac{dP}{dt}\right]$ are not
necessarily zero, time derivatives of powers of $X$ and $P$ must be taken term
by term.  Through the Laurent series, this can then be used to find the time
derivative of arbitrary functions of $X$ and $P$.

Before we can define the time derivative of the Laurent series, we must first
define the time derivative of $X^{-1}$, which can be found through
\begin{eqnarray}
\nonumber
\frac{d}{dt} X^{-1} &=& \frac{d}{dt}\left(X^{-1}XX^{-1}\right)\\
&=&2\frac{d}{dt}X^{-1} + X^{-1}\frac{P}{m}X^{-1}
\end{eqnarray}
which implies
\begin{equation}
\frac{d}{dt} X^{-1} = -X^{-1}\frac{P}{m}X^{-1}
\end{equation}
and by the same argument
\begin{equation}
\frac{d}{dt} X^{-n} = -X^{-n}\left(\frac{d}{dt}X^n\right)X^{-n}
\label{Inversedt}
\end{equation}

The time derivative of $X^n$ can be found term by term as:
\begin{eqnarray}
\nonumber
\frac{d}{dt}X^n &=&
\frac{dX}{dt}X^{n-1}+X\frac{dX}{dt}X^{n-2}+\ldots+X^{n-1}\frac{dX}{dt}\\
&=& \frac{P}{m}X^{n-1}+X\frac{P}{m}X^{n-2}+\ldots+X^{n-1}\frac{P}{m}
\end{eqnarray}

Commuting all of the $P$'s to left, this equation becomes
\begin{equation}
\frac{d}{dt}X^n=\frac{1}{m}\left(nPX^{n-1} +
\sum_{j=1}^{n-1}\left[X^j,PX^{n-j}\right]\right)
\end{equation}
and using the fact that $nX^{n-1} = -i\left[X^n,P\right]$, we can write
this as
\begin{equation}
\frac{d}{dt}X^n=\frac{1}{m}\left(-iP\left[X^n,P\right] +
\sum_{j=1}^{n-1}\left[X^j,PX^{n-j}\right]\right)
\label{XDeriv1}
\end{equation}

If instead, we commute all the $P$'s to the right, we get
\begin{equation}
\frac{d}{dt}X^n=\frac{1}{m}\left(-i\left[X^n,P\right]P -
\sum_{j=1}^{n-1}\left[X^j,PX^{n-j}\right]\right)
\label{XDeriv2}
\end{equation}
where the minus sign on the second commutator is picked up because we have
commuted the $P$'s to the opposite side.

Since both equations \ref{XDeriv1} and \ref{XDeriv2} are equal to
$\frac{d}{dt}X^n$, the average of the two of them is still equal to
$\frac{d}{dt}X^n$, and we can write
\begin{equation}
\frac{d}{dt}X^n =
\frac{-i}{2m}\left(P\left[X^n,P\right]+\left[X^n,P\right]P\right) =
i\left[\frac{P^2}{2m},X^n\right]
\label{XDerivAvg}
\end{equation}
for positive values of n.

For inverse powers of $X$, we can now rewrite equation \ref{Inversedt} as
\begin{equation}
\frac{d}{dt}X^{-n} = -X^{-n}i\left[\frac{P^2}{2m},X^n\right]X^{-n} =
i\left[\frac{P^2}{2m},X^{-n}\right]
\end{equation}
and so, for an arbitrary function of $X$, via the Laurent expansion
\begin{equation}
\frac{d}{dt}O(X) = i\left[\frac{P^2}{2m},O(X)\right]
\end{equation} 

There is another way of arriving at the same result that we found above which is
useful when $\frac{d}{dt}X$ is a more general function of $P$, as in the
relativistic case.  For a velocity that is an arbitrary function of
momentum $V(P) = \frac{d}{dt}X$, we can make the substitution
\begin{equation}
V(P) = -i\left[X,\int V(P)dP\right]
\label{HPart1}
\end{equation}
that is, $V$ is the derivative of the integral of $V(P)$ with respect to $P$. 
The time derivative of $X^n$ becomes
\begin{eqnarray}
\nonumber
\frac{dX^n}{dt} &=& i\left[\int V(P)dP,X\right]X^{n-1}+iX\left[\int
V(P)dP,X\right]X^{n-1}+\ldots\\
&=& i\left[\int V(P)dP,X^n\right]
\label{VIntDeriv}
\end{eqnarray}
and the time derivative of $O(X)$ is
\begin{equation}
\frac{d}{dt}O(X) = i\left[\int V(P)dP, O(X)\right]
\end{equation}
It is easy to see, that for the Newtonian velocity/momentum relationship,
this returns the usual $\frac{P^2}{2m}$ commutator.

This method can be employed again for finding the time derivative of $P^n$. 
Since the force, $F$, can be an arbitrary function of $X$, there is no simple
algebraic way of taking the time derivative as in equation \ref{XDerivAvg}. 
But, by making the substitution
\begin{equation}
F(X) = i\left[P,\int F dX\right]
\label{HPart2}
\end{equation}
we can find the time derivative of $P^n$ by the same method that we used to get
equation \ref{VIntDeriv}.  We see then, that
\begin{equation}
\frac{d}{dt}O(P) = -i\left[\int F dX, O(P)\right]
\end{equation}

A function of $X$ and $P$, $O(X,P)$ can be Laurent expanded as
\begin{equation}
O(X,P) =
\sum_{-\infty}^{\infty}C_{nmjk\ldots}
X^nP^mX^jP^k\ldots
\label{XPLaurent}
\end{equation}
with an arbitrary number of alternating powers of $X$ and $P$ where the indexed
coefficient is a constant and the summation is taken over each independent
power n, m, j, k, etc.  Commuting this series with $-F(X)$ and $V(P)$ gives us
\begin{eqnarray}
-\sum_{-\infty}^{\infty}C_{nmjk\ldots}\left(X^n\left[F(X),P^m\right]X^jP^k\ldots
+ XnP^mX^j\left[F(X),P^k\right]\ldots + \ldots\right)\\
\sum_{-\infty}^{\infty}C_{nmjk\ldots}\left(\left[V(P),X^n\right]P^mX^jP^k\ldots
+ X^nP^m\left[V(P),X^j\right]P^k\ldots + \ldots\right)
\end{eqnarray}
and it is clear that the sum of these two series is the full time derivative of
$O(X,P)$, differentiated term-by-term, via the chain rule. Thus, for an
arbitrary function $O(X,P)$, the time derivative can be written as
\begin{equation}
\frac{d}{dt}O(X,P) = i\left[\int V(P)dP - \int F(X)dX, O(X,P)\right]
\label{GeneralHEquation}
\end{equation}
or specifically, in Newtonian mechanics
\begin{equation}
\frac{d}{dt}O(X,P) = i\left[\frac{P^2}{2m} - \int F(X)dX, O(X,P)\right]
\label{HEquation}
\end{equation}
which is exactly the Heisenberg equation of motion.  Equation \ref{HEquation}
provides a complete description of Heisenberg picture quantum mechanics and can
be used to solve for the time propagator $U(t) = \exp\left\{-iHt\right\}$.

It is no coincidence that the integrals $\int F(X)dX$ and $\int V(P)dP$ that
appear in equation \ref{GenearlHEquation} when added together produce the
Hamiltonian.
From Hamiltons equations:
\begin{eqnarray}
\frac{\partial H}{\partial X} &=& -\dot{P}\\
\frac{\partial H}{\partial P} &=& \dot{X}
\end{eqnarray}
and thus, for a Hamiltonian that is separable into $H(X,P) = H(X) + H(P)$ we can
write
\begin{equation}
H(X,P) = \int \dot{X}dP - \int \dot{P}dX
\end{equation}

Equation \ref{HEquation} is the quantum equivalent of
\begin{equation}
\frac{d}{dt}O(X,P) = \frac{\partial O}{\partial X}\frac{P}{m} +
\frac{\partial O}{\partial P}F
\end{equation}
but in a way that respects the matrix properties of the $X$ and $P$ operators.

By taking the derivative in this manner, we have reproduced Heisenberg picture
quantum mechanics, that is, the fact that the time derivative of an operator is
proportional to its commutator with the Hamiltonian.  We have done so without
resorting to energy, conserved quantities, or even the term Hamiltonian itself. 
Instead, the integrals of force and velocity appeared as a way of simplifying
the commutators that arose in our calculations.

This derivation, however, ultimately results in the use of the Hamiltonian,
whether referred to as such or not, and does not clearly underscore the fact
that the quantum operators for position and momentum evolve in time in a way
that is very similar to their classical counterparts under Newton's laws. After
all, Newton's laws do not make use of any analogous method of taking partial
derivatives and typically only involve $X$ and its derivatives, rather than
general functions of $X$ and $P$.  In the next section, we will explore a
formulation of the propagator that highlights the Newtonian-like dynamics of the
operator $X(t)$.

\section{The Propagator from the Newtonian dynamics of X(t)}
Just as in classical mechanics, in quantum mechanics, $X(t)$ can be written as
\begin{equation}
X(t) = X_0 + \frac{1}{m}P_0t + \frac{1}{2m}F_0t^2 +
\frac{1}{6m}\frac{dF}{dt}t^3+\ldots
\label{XT1}
\end{equation}
the difference being that $X_0$ and $P_0$ are matrices that obey the canonical
commutation relation.

For simplicity, we can rewrite equation \ref{XT1} as
\begin{equation}
X(t) = X_0 + \frac{1}{m}\int_0^t P(t)dt
\end{equation}
where $P(t)$ is a matrix with a complicated time dependence determined by the
force, $F(X)$.

At any time, t, there is a vector $|X_a;t\rangle$ that is an eigenvector
of $X(t)$ with eigenvalue $x_a$, such that
\begin{equation}
X(t)|X_a;t\rangle =\left(X_0 + \frac{1}{m}\int_0^t P(t)dt\right)|X_a;t\rangle=
x_a|X_a;t\rangle
\end{equation}

At $t=0$, this eigenvector is the Dirac delta function
$|X_a;0\rangle=\delta(X-X_a)$, but at a later time $t$ is given by
\begin{equation}
|X_a;t\rangle = U^\dagger(t)|X_a;0\rangle
\end{equation}
since $X(t)$ evolves according to $X(t) = U^\dagger(t)X_0U(t)$.

We can take the expectation value of $X(t)$ with two different eigenvectors at
two different times to find $\langle X_b;0|X(t)| X_a;t\rangle$ and
$\langle X_b;t|X(t)| X_a;0\rangle$ which gives us
\begin{eqnarray}
\label{Prop1}
\langle X_b|\left(X_0+\frac{1}{m}\int_0^t P(t)dt\right) U^\dagger(t)| X_a\rangle
&=& x_a\langle X_b|U^\dagger(t)|X_a\rangle\\
\label{Prop2}
\langle X_b|U(t) \left(X_0+\frac{1}{m}\int_0^t P(t)dt\right)| X_a\rangle &=&
x_b\langle X_b|U(t)|X_a\rangle
\end{eqnarray}
where $|X_a\rangle$ and $|X_b\rangle$ are taken to be the eigenvectors at $t=0$.

It is worth noting that if we allow $X_0$ to act on $\langle X_b|$ of equation
\ref{Prop1}, we can write
\begin{equation}
\langle X_b|\int_0^t P(t)dt U^\dagger|X_a\rangle = (x_b-x_a)\langle
X_b|U^\dagger(t)|X_a\rangle
\end{equation}

The left hand side of the equation contains the integral of momentum with
respect to time, and the right hand contains the displacement $\Delta x = x_b -
x_a$.  In other words, we have written the quantum analog of the
classical equation $\Delta X = \int_0^t P dt$.

In principle, finding the propagator $\langle X_b|U(t)|X_a\rangle$ amounts to
finding the solution to equations \ref{Prop1} and \ref{Prop2}.  In practice,
this can be difficult, although there are at least three cases that admit an
exact solution. A complete differential equation for the propagator can be
written with this method if and only if an exact solution for the time dependent
operators $X(t)$ and $P(t)$ can be found. In the three cases described in this
paper, the time derivatives of $X(t)$ and $P(t)$ at $t=0$ are at most linear in
$X_0$ or $P_0$. Because of this, repeated differentiation will not cause
mixtures of alternating powers of $X_0$ and $P_0$, the Taylor series in time can
be written to infinite order, and the exact operators plugged into equations
\ref{Prop1} and \ref{Prop2}.

Since equations \ref{Prop1} and \ref{Prop2} do not include the time derivative
of $U(t)$, there is the possibility that our solution could differ from the true
propagator either by a purely time dependent factor $A(t)$ or by an additional
purely time dependent term $g(t)$ that needs to be added to it. The fact that
$U(t)$ is unitary, precludes the possibility that a purely time dependent
function could be added to our solution, since this would change the magnitude
of $U(t)$ with time, and thus, $g(t)$ must equal zero.

$A(t)$ can be determined by the criterion that $U(t) = \delta(x_b-x_a)$ at
$t=0$.  Any additional time dependent factor cannot affect the amplitude of
$U(t)$, again because of unitarity.  Although this does not rule out time
dependent phase factors, such a factor would be the equivalent of at most
shifting the potential by a time dependent, real function $f(t)$ that is
constant over all space.  Such a time dependent change in phase cannot affect
any measurable properties of the system. In other words, the requirement that
$U(t)$ be unitary restricts the possible solutions to physically equivalent
expressions.

\subsection{The Free Particle}

If $F(X)$ is zero everywhere, $\int P(t) dt$ becomes $P_0t$.   It is convenient
to let $X(t)$ act to the left in equation \ref{Prop1} and to the right in
equation \ref{Prop2}.  The operator $P_0$ can then be defined by its action on
$\langle X_b|$ and $|X_a\rangle$ as
\begin{eqnarray}
\nonumber
P|X_a\rangle &=& \int_{-\infty}^{\infty} P_0|P_0\rangle\langle P_0|X_a\rangle
dP_0 = \int_{-\infty}^{\infty} P_0|P_0\rangle e^{-iP_0x_a}dP_0 =
\int_{-\infty}^{\infty} i\frac{\partial}{\partial x_a}|P_0\rangle
e^{-iP_0x_a}dP_0\\
&=& i\frac{\partial}{\partial x_a}|X_a\rangle
\end{eqnarray}
and through the same procedure
\begin{eqnarray}
\langle X_b|P_0 = -i\frac{\partial}{\partial X_b}\langle X_b|
\end{eqnarray}

Equations \ref{Prop1} and \ref{Prop2} then become
\begin{eqnarray}
\label{Prop1Free}
x_b\langle X_b|U^\dagger(t) |X_a\rangle-\frac{it}{m}\frac{\partial}{\partial
x_b}\langle X_b| U^\dagger(t)| X_a\rangle &=& x_a\langle
X_b|U^\dagger(t)|X_a\rangle\\
\label{Prop2Free}
x_a\langle X_b|U(t)| X_a\rangle + \frac{it}{m}\frac{\partial}{\partial
x_a}\langle X_b|U(t)| X_a\rangle &=& x_b\langle X_b|U(t)|X_a\rangle
\end{eqnarray}
where the derivative operator has different signs in \ref{Prop1Free} and
\ref{Prop2Free} because it is acting to the left and to the right, respectively.

Relabeling $\langle X_b|U(t)|X_a\rangle$ as $U(x_b,x_a,t)$, we can turn equation
\ref{Prop1Free} into the integral equation
\begin{equation}
\int \frac{dU^\dagger(x_b,x_a,t)}{U^\dagger(x_b,x_a,t)} =
\frac{im}{t}\int(x_a-x_b)dx_b
\end{equation}
which has the solution
\begin{equation}
U^\dagger(x_b,x_a,t) = A^\dagger(t)\exp\left\{-im\left(\frac{\frac{1}{2}x_b^2 -
x_bx_a + f(x_a)}{t}\right)\right\}
\label{FreeSol1}
\end{equation}

By the same method, the solution to equation \ref{Prop2Free} is
\begin{equation}
U(x_b,x_a,t) = A(t)\exp\left\{im\left(\frac{\frac{1}{2}x_a^2 - x_bx_a +
f(x_b)}{t}\right)\right\}
\label{FreeSol2}
\end{equation}

The solutions of equations \ref{FreeSol1} and \ref{FreeSol2} set $f(x_a) =
\frac{1}{2}x_a^2$ and $f(x_b)=\frac{1}{2}x_b^2$.  Furthermore, the boundary
condition $U(x_b,x_a,0) = \delta(x_a-x_b)$ determines $A(t)$, so that the
propagator is equal to
\begin{equation}
U(x_b,x_a,t) = \left(\frac{m}{2\pi
it}\right)^{1/2}\exp\left\{\frac{im}{2t}\left(x_b-x_a\right)^2\right\}
\end{equation}
which correctly matches the known solution.

\subsection{The Harmonic Oscillator Propagator}
To solve the propagator for the force $F(x)=-\omega^2mX$, we can Taylor expand
$X(t)$ to get:
\begin{eqnarray}
\nonumber
X(t) &=& X_0 + \frac{P_0}{m}t - \frac{\omega^2}{2}X_0t^2 -
\frac{\omega^2}{6}\frac{P_0}{m}t^3 +\ldots\\
&=& X_0\cos(\omega t) + \frac{P_0}{m\omega^2}\sin(\omega t)
\end{eqnarray}

Equations \ref{Prop1} and \ref{Prop2} then become
\begin{eqnarray}
\label{Prop1Harmonic}
\langle X_b|\left(X_0\cos(\omega t) +
\frac{P_0}{m\omega^2}\sin(\omega t)\right)U^\dagger(t) | X_a\rangle &=&
x_a\langle X_b|U^\dagger|X_a\rangle\\
\label{Prop2Harmonic}
\langle X_b|U(t) \left(X_0\cos(\omega t) +
\frac{P_0}{m\omega^2}\sin(\omega t)\right)| X_a\rangle &=& x_b\langle
X_b|U|X_a\rangle
\end{eqnarray}

Equation \ref{Prop1Harmonic} can be turned into an integral equation, as with
the free particle, yielding
\begin{equation}
\int \frac{dU^\dagger(x_b,x_a,t)}{U^\dagger(x_b,x_a,t)}=im\omega\int
\frac{x_a - x_b\cos(\omega t)}{\sin(\omega t)}dx_b
\end{equation}

Combined with the solution to equation \ref{Prop2Harmonic} and, once again, the
condition that \newline$U(x_b,x_a,0)\nobreak=\nobreak\delta(x_b\nobreak-\nobreak
x_a)$, we get
\begin{equation}
U(x_b,x_a,t) = \left(\frac{m\omega}{2\pi i\sin(\omega
t)}\right)^{1/2}\exp\left\{\-\frac{m\omega((x_b^2+x_a^2)\cos(\omega
t)-2x_bx_a)}{2i\sin(\omega t)}\right\}
\end{equation}
which, again, matches the known result.

\subsection{The Constant Force Propagator}
If a constant force is applied to a particle, $F(t) = F_0$, corresponding
to the potential $\mathcal{U}(x) = -F_0 X$, then $X(t)$ and $P(t)$ can be solved
exactly and are
\begin{eqnarray}
P(t) &=& P_0 + F_0 t\\
X(t) &=& X_0 + \frac{P_0}{m}t + \frac{1}{2m}F_0t^2
\end{eqnarray}

This adds only a small amount of complexity beyond the free particle case. 
Equations \ref{Prop1} and \ref{Prop2} become
\begin{eqnarray}
\label{Prop1Const}
(x_b+\frac{1}{2m}F_0t^2)\langle X_b|U^\dagger(t)
|X_a\rangle-\frac{it}{m}\frac{\partial}{\partial x_b}\langle X_b| U^\dagger(t)| X_a\rangle &=& x_a\langle X_b|U^\dagger|X_a\rangle\\
\label{Prop2Const}
(x_a+\frac{1}{2m}F_0t^2)\langle X_b|U(t)| X_a\rangle +
\frac{it}{m}\frac{\partial}{\partial x_a}\langle X_b|U(t)| X_a\rangle &=& x_b\langle X_b|U|X_a\rangle
\end{eqnarray}

The solution to equations \ref{Prop1Const} and \ref{Prop2Const}, using the same
integral method as in the free particle case, is
\begin{equation}
U(x_b,x_a,t) = \left(\frac{m}{2\pi
it}\right)^{1/2}\exp\left\{\frac{im}{2t}\left(\left(x_b-x_a\right)^2 +
\frac{1}{m}F_0t^2(x_b+x_a)\right)\right\}
\end{equation}
where the coefficient out front is set by the same delta function boundary
condition.  Again, this matches the known propagator\cite{LinearProp,
LinearProp2} up to a phase factor that is constant over all space and the result
is achieved in a very simple fashion, since $X(t)$ is easily solvable for a
constant force.

\section{Connection to the Path Integral}

Although the propagator was only solved for three particular cases where the
time dependence of $X(t)$ and $P(t)$ could be solved exactly, this technique
is, in theory, applicable to particles under the influence of any arbitrary
force $F(X)$.  Although the exact differential equation for the propagator can
only be written when there is an analytic solution to the time dependence
of $X(t)$, it is always possible to write an approximate solution to the
propagator over a small time interval.  We will show that by piecing together
propagators over small intervals, we can use this technique to reproduce the
Feynman path integral formula, much in the same way as it can be derived
starting with the Hamiltonian formalism.

As stated in equation \ref{XT1}, $X(t)$ can be Taylor expanded in terms of
$P_0$, $F_0$, and further time derivatives.  If we keep only the terms to second
order in time, for a small time interval, $\Delta t$, we get
\begin{equation}
X(t)\approx X_0 + \frac{1}{m}P_0\Delta t + \frac{1}{2m}F(X_0)\Delta t^2
\end{equation}

Using this approximate $X(t)$, we can write the differential equation for the
propagator over a small time interval $U(\Delta t)$ as
\begin{eqnarray}
\label{ApproxProp1}
(x_b+\frac{1}{2m}F(x_b)\Delta t^2)\langle X_b|U^\dagger(\Delta t)
|X_a\rangle-\frac{i\Delta t}{m}\frac{\partial}{\partial x_b}\langle X_b|
U^\dagger(\Delta t)| X_a\rangle &=& x_a\langle X_b|U^\dagger(\Delta
t)|X_a\rangle\\
\label{ApproxProp2}
(x_a+\frac{1}{2m}F(x_a)\Delta t^2)\langle X_b|U(\Delta t)| X_a\rangle +
\frac{i\Delta t}{m}\frac{\partial}{\partial x_a}\langle X_b|U(\Delta t)|
X_a\rangle &=& x_b\langle X_b|U(\Delta t)|X_a\rangle
\end{eqnarray}

Equation \ref{ApproxProp1} becomes the integral equation
\begin{equation}
\int \frac{dU^\dagger(\Delta t)}{U^\dagger(\Delta t)}=\frac{im}{\Delta t}\int
\left(x_a - x_b -\frac{1}{2m}F(x_b)\Delta t^2\right) dx_b
\end{equation}
which has the solution
\begin{eqnarray}
U^\dagger(\Delta t) = A^\dagger(\Delta
t)\exp\left\{\frac{im}{\Delta t}\left(x_ax_b -\frac{1}{2}x_a^2 -
\frac{1}{2m}\int F(x_b)dx_b\Delta t^2 +f(x_a)\right)\right\}
\end{eqnarray}
where $A(\Delta t)$ is defined, as in the previous section, to be a factor that
will set the boundary condition that $U(t)$ is a delta function at $t=0$.  Solving
equation \ref{ApproxProp2} in a similar manner fixes $f(x_a)$ and we find that
the propagator is
\begin{equation}
U(x_b,x_a,\Delta t) =A(\Delta
t)\exp\left\{\frac{im}{2\Delta t}\left(\left(x_b-x_a\right)^2 +
\frac{1}{m}\left(\int F(x_a) dx_a +\int F(x_b) dx_b\right) \Delta t^2\right)\right\}
\label{DTProp}
\end{equation}

Since equation \ref{DTProp} is only valid in the limit of small $\Delta t$, to
calculate a propagator that spans a larger time period, we can subdivide
the time interval into N smaller steps and string together several propagators
over small $\Delta t$. Since only the endpoints ($x_1$ and $x_N$, corresponding
to the initial and final locations) are fixed, we must integrate over all
intermediate locations, and we get
\begin{eqnarray}
\nonumber
\langle x_N|U(t)|x_1\rangle &=&A(t)\int dx_2\ldots dx_{N-1} \langle x_N|U(\Delta
t)|x_{N-1}\rangle\langle x_{N-1}|U(\Delta t)|x_{N-2}\rangle\ldots\langle x_2|U(\Delta t)|x_1\rangle\\
&=& A(t) \int dx_2\ldots dx_{N-1}\prod_{i=2}^N
e^{\frac{im}{2\Delta t}\left(\left(x_i-x_{i-1}\right)^2 +\frac{1}{m}\left(\int
F(x_{i-1}) dx_{i-1}\linebreak +\int F(x_{i}) dx_{i}\right) \Delta t^2\right)}
\label{PathProp1}
\end{eqnarray}
where all of the factors $A(\Delta t)$ have been combined into a single factor,
$A(t)$ that enforces the boundary condition at $t=0$.

Noting that $x_i-x_{i-1} = \frac{1}{m}p_{i+\frac{1}{2}}\Delta t$, where
$p_{i+\frac{1}{2}}$ represents the average momentum on the interval between $x_{i-1}$ and $x_i$, and that $\int
F(x)dx = -\mathcal{U}(x)$, we can rewrite \ref{PathProp1} as
\begin{equation}
U(x_b,x_a,t)=A(t) \int dx_2\ldots dx_{N-1}\prod_{i=2}^N
e^{\frac{i}{2m}p_{i+\frac{1}{2}}^2\Delta t -\frac{i}{2}\left(\mathcal{U}(x_i) +
\mathcal{U}(x_{i-1})\right) \Delta t}
\end{equation}
The $\frac{1}{2}\left(\mathcal{U}(x_i) + \mathcal{U}(x_{i-1})\right)$ in the
exponent is approximately average potential between $x_{i-1}$ and $x_i$.  This is true
since we are considering $\Delta t$ to be a very small time interval and will
eventually take the limit as $\Delta t$ goes to zero.  We can then make the
substitution that $\frac{1}{2}\left(\mathcal{U}(x_i) +
\mathcal{U}(x_{i-1})\right)=\mathcal{U}(x_{i+\frac{1}{2}})$. the The term that
appears in the exponent, $\frac{1}{2m}p_{i+\frac{1}{2}}^2 -
\mathcal{U}(x_{i+\frac{1}{2}})$, is the Lagrangian, $\mathcal{L}$. Furthermore,
the product of exponentials can be turned into a sum of exponents, leaving us
with
\begin{equation}
U(x_b,x_a,t)=A(t) \int dx_2\ldots dx_{N-1}e^{
i\sum_{i=1}^{N-1}\mathcal{L}(x_{i+\frac{1}{2}},p_{i+\frac{1}{2}}) \Delta t}
\end{equation}

In the limit that we subdivide into an infinite number of infinitesimal
intervals, each spanning an infinitesimal $\Delta t$ we arrive at our final
expression for the propagator
\begin{equation}
U(x_b,x_a,t)=A(t) \int Dx(t) e^{
i\int \mathcal{L}(x(t),p(t)) dt}
\end{equation}
where the capital D refers to a sum over all paths $x(t)$.  This is exactly the
expression derived by Feynman for obtaining the propagator with the path
integral method\cite{PathIntegral}.

\section{Concluding Remarks}
There are still issues that are difficult to address in a Newtonian formulation
of physics, such as the fact that the momentum operator
$P=-i\frac{\partial}{\partial X}$ is the canonical momentum, rather than
$mV$. This can necessitate, as in the case of the Aharanov-Bohm problem, the
addition of a term whose interpretation is unclear in Newtonian mechanics to
produce the standard Newtonian momentum.

The fact that the integral of force that appears in equation \ref{HEquation} is
an indefinite integral is also confusing in the case of a delta function force,
which corresponds to a discontinuous, step function potential.  Without the
motivation of a well defined potential energy function, it is difficult to see
why the integral at every point must be defined in such a way that there is a
step at the location of the force, although it may be possible to hand wave an
argument based on the non-locality of momentum states that the force acts on.

Despite these interpretational difficulties for certain classes of problems,
this formulation of quantum mechanics provides a key connection between
Hamiltonian, Lagrangian, and Newtonian formulations of physics, even in the
quantum regime.  Especially for students who are new to Hamiltonian and
Lagrangian mechanics, it can be used to form a bridge to facilitate the
transition from their old way of thinking about physics to the new, and often
seemingly bizarre quantum regime.

\bibliography{Citations}{}

\begin{thebibliography}{11}%
\makeatletter
\providecommand \@ifxundefined [1]{%
 \@ifx{#1\undefined}
}%
\providecommand \@ifnum [1]{%
 \ifnum #1\expandafter \@firstoftwo
 \else \expandafter \@secondoftwo
 \fi
}%
\providecommand \@ifx [1]{%
 \ifx #1\expandafter \@firstoftwo
 \else \expandafter \@secondoftwo
 \fi
}%
\providecommand \natexlab [1]{#1}%
\providecommand \enquote  [1]{``#1''}%
\providecommand \bibnamefont  [1]{#1}%
\providecommand \bibfnamefont [1]{#1}%
\providecommand \citenamefont [1]{#1}%
\providecommand \href@noop [0]{\@secondoftwo}%
\providecommand \href [0]{\begingroup \@sanitize@url \@href}%
\providecommand \@href[1]{\@@startlink{#1}\@@href}%
\providecommand \@@href[1]{\endgroup#1\@@endlink}%
\providecommand \@sanitize@url [0]{\catcode `\\12\catcode `\$12\catcode
  `\&12\catcode `\#12\catcode `\^12\catcode `\_12\catcode `\%12\relax}%
\providecommand \@@startlink[1]{}%
\providecommand \@@endlink[0]{}%
\providecommand \url  [0]{\begingroup\@sanitize@url \@url }%
\providecommand \@url [1]{\endgroup\@href {#1}{\urlprefix }}%
\providecommand \urlprefix  [0]{URL }%
\providecommand \Eprint [0]{\href }%
\providecommand \doibase [0]{http://dx.doi.org/}%
\providecommand \selectlanguage [0]{\@gobble}%
\providecommand \bibinfo  [0]{\@secondoftwo}%
\providecommand \bibfield  [0]{\@secondoftwo}%
\providecommand \translation [1]{[#1]}%
\providecommand \BibitemOpen [0]{}%
\providecommand \bibitemStop [0]{}%
\providecommand \bibitemNoStop [0]{.\EOS\space}%
\providecommand \EOS [0]{\spacefactor3000\relax}%
\providecommand \BibitemShut  [1]{\csname bibitem#1\endcsname}%
\let\auto@bib@innerbib\@empty
%</preamble>
\bibitem [{\citenamefont {Ehrenfest}(1927)}]{ehrenfest}%
  \BibitemOpen
  \bibfield  {author} {\bibinfo {author} {\bibfnamefont {P.}~\bibnamefont
  {Ehrenfest}},\ }\href@noop {} {\bibfield  {journal} {\bibinfo  {journal}
  {Zeitschrift f{\"u}r Physik}\ }\textbf {\bibinfo {volume} {45}},\ \bibinfo
  {pages} {455} (\bibinfo {year} {1927})}\BibitemShut {NoStop}%
\bibitem [{\citenamefont {Weinberg}(2012)}]{weinbergQM}%
  \BibitemOpen
  \bibfield  {author} {\bibinfo {author} {\bibfnamefont {S.}~\bibnamefont
  {Weinberg}},\ }\href {http://books.google.com/books?id=WfTq2W\_LBlEC} {\emph
  {\bibinfo {title} {Lectures on Quantum Mechanics}}}\ (\bibinfo  {publisher}
  {Cambridge University Press},\ \bibinfo {year} {2012})\BibitemShut {NoStop}%
\bibitem [{\citenamefont {Griffiths}(2005)}]{griffithsQM}%
  \BibitemOpen
  \bibfield  {author} {\bibinfo {author} {\bibfnamefont {D.}~\bibnamefont
  {Griffiths}},\ }\href@noop {} {\emph {\bibinfo {title} {Introduction to
  Quantum Mechanics}}}\ (\bibinfo  {publisher} {Pearson Prentice Hall},\
  \bibinfo {year} {2005})\BibitemShut {NoStop}%
\bibitem [{\citenamefont {Sakurai}(1993)}]{SakuraiQM}%
  \BibitemOpen
  \bibfield  {author} {\bibinfo {author} {\bibfnamefont {J.~J.}\ \bibnamefont
  {Sakurai}},\ }\href {http://www.worldcat.org/isbn/0201539292} {\emph
  {\bibinfo {title} {{Modern Quantum Mechanics (Revised Edition)}}}},\ \bibinfo
  {edition} {1st}\ ed.\ (\bibinfo  {publisher} {Addison Wesley},\ \bibinfo
  {year} {1993})\BibitemShut {NoStop}%
\bibitem [{\citenamefont {Aitchison}\ \emph {et~al.}(2004)\citenamefont
  {Aitchison}, \citenamefont {MacManus},\ and\ \citenamefont
  {Snyder}}]{AJPHeisenberg}%
  \BibitemOpen
  \bibfield  {author} {\bibinfo {author} {\bibfnamefont {I.~J.~R.}\
  \bibnamefont {Aitchison}}, \bibinfo {author} {\bibfnamefont {D.~A.}\
  \bibnamefont {MacManus}}, \ and\ \bibinfo {author} {\bibfnamefont {T.~M.}\
  \bibnamefont {Snyder}},\ }\href {\doibase
  http://dx.doi.org/10.1119/1.1775243} {\bibfield  {journal} {\bibinfo
  {journal} {American Journal of Physics}\ }\textbf {\bibinfo {volume} {72}},\
  \bibinfo {pages} {1370} (\bibinfo {year} {2004})}\BibitemShut {NoStop}%
\bibitem [{\citenamefont {Van~Vleck}(1928)}]{VVCorrespondence}%
  \BibitemOpen
  \bibfield  {author} {\bibinfo {author} {\bibfnamefont {J.~H.}\ \bibnamefont
  {Van~Vleck}},\ }\href@noop {} {\bibfield  {journal} {\bibinfo  {journal}
  {Proceedings of the National Academy of Sciences of the United States of
  America}\ }\textbf {\bibinfo {volume} {14}},\ \bibinfo {pages} {178}
  (\bibinfo {year} {1928})}\BibitemShut {NoStop}%
\bibitem [{\citenamefont {Feynman}(1948)}]{PathIntegral}%
  \BibitemOpen
  \bibfield  {author} {\bibinfo {author} {\bibfnamefont {R.~P.}\ \bibnamefont
  {Feynman}},\ }\href {\doibase 10.1103/RevModPhys.20.367} {\bibfield
  {journal} {\bibinfo  {journal} {Rev. Mod. Phys.}\ }\textbf {\bibinfo {volume}
  {20}},\ \bibinfo {pages} {367} (\bibinfo {year} {1948})}\BibitemShut
  {NoStop}%
\bibitem [{\citenamefont {de~la Peña-Auerbach}(1969)}]{StochasticQM}%
  \BibitemOpen
  \bibfield  {author} {\bibinfo {author} {\bibfnamefont {L.}~\bibnamefont
  {de~la Peña-Auerbach}},\ }\href {\doibase
  http://dx.doi.org/10.1063/1.1665009} {\bibfield  {journal} {\bibinfo
  {journal} {Journal of Mathematical Physics}\ }\textbf {\bibinfo {volume}
  {10}},\ \bibinfo {pages} {1620} (\bibinfo {year} {1969})}\BibitemShut
  {NoStop}%
\bibitem [{\citenamefont {Nelson}(1966)}]{SchrodingerFromNewton}%
  \BibitemOpen
  \bibfield  {author} {\bibinfo {author} {\bibfnamefont {E.}~\bibnamefont
  {Nelson}},\ }\href {\doibase 10.1103/PhysRev.150.1079} {\bibfield  {journal}
  {\bibinfo  {journal} {Phys. Rev.}\ }\textbf {\bibinfo {volume} {150}},\
  \bibinfo {pages} {1079} (\bibinfo {year} {1966})}\BibitemShut {NoStop}%
\bibitem [{\citenamefont {Nassar}\ \emph {et~al.}(2002)\citenamefont {Nassar},
  \citenamefont {Bassalo}, \citenamefont {Alencar}, \citenamefont {Lopes},
  \citenamefont {Oliveira},\ and\ \citenamefont {Cattani}}]{LinearProp}%
  \BibitemOpen
  \bibfield  {author} {\bibinfo {author} {\bibfnamefont {A.~A.~B.}\
  \bibnamefont {Nassar}}, \bibinfo {author} {\bibfnamefont {J.~A. M.~F.}\
  \bibnamefont {Bassalo}}, \bibinfo {author} {\bibfnamefont {P.~T.~S.}\
  \bibnamefont {Alencar}}, \bibinfo {author} {\bibfnamefont {J.~A. L.~M.}\
  \bibnamefont {Lopes}}, \bibinfo {author} {\bibfnamefont {J.~A. I. F.~d.}\
  \bibnamefont {Oliveira}}, \ and\ \bibinfo {author} {\bibfnamefont {M.~S.~D.}\
  \bibnamefont {Cattani}},\ }\href
  {http://www.scielo.br/scielo.php?script=sci_arttext&pid=S0103-97332002000400024&nrm=iso}
  {\bibfield  {journal} {\bibinfo  {journal} {{Brazilian Journal of Physics}}\
  }\textbf {\bibinfo {volume} {32}},\ \bibinfo {pages} {812 } (\bibinfo {year}
  {2002})}\BibitemShut {NoStop}%
\bibitem [{\citenamefont {Holstein}(1997)}]{LinearProp2}%
  \BibitemOpen
  \bibfield  {author} {\bibinfo {author} {\bibfnamefont {B.~R.}\ \bibnamefont
  {Holstein}},\ }\href {\doibase http://dx.doi.org/10.1119/1.18550} {\bibfield
  {journal} {\bibinfo  {journal} {American Journal of Physics}\ }\textbf
  {\bibinfo {volume} {65}},\ \bibinfo {pages} {414} (\bibinfo {year}
  {1997})}\BibitemShut {NoStop}%
\end{thebibliography}%
\end{document}